\documentclass[pra,10pt,showpacs,eqsecnum,amsmath,onecolumn,floatfix]{revtex4}
\usepackage{amsmath}
\usepackage{amsfonts}
\usepackage{graphicx}

\newcommand{\diff}[2]{\frac{d #1}{d #2}}

\newcommand{\intall}{\int_{-\infty}^{\infty}}
\newcommand{\ket}[1]{|#1\rangle}

\newcommand{\bra}[1]{\langle#1|}

\newcommand{\Avg}[1]{\left\langle#1\right\rangle}

\newcommand{\abs}[1]{\left|#1\right|}
\newcommand{\bk}[1]{\left(#1\right)}
\newcommand{\Bk}[1]{\left[#1\right]}

\newcommand{\trace}{\operatorname{tr}}

\begin{document}
\title{Cavity quantum electro-optics. II. Input-output relations
  between traveling optical and microwave fields}

\author{Mankei Tsang}

\email{mankei@unm.edu}
\affiliation{Department of Electrical and Computer Engineering,
  National University of Singapore, 4 Engineering Drive 3, Singapore
  117576}

\affiliation{Department of Physics, National University of Singapore,
  2 Science Drive 3, Singapore 117542}

\affiliation{
Center for Quantum Information and Control,
University of New Mexico, MSC07--4220, Albuquerque, New Mexico
87131-0001, USA}








\date{\today}

\begin{abstract}
  In the previous paper [M.~Tsang, \pra \textbf{81}, 063837 (2010)], I
  proposed a quantum model of a cavity electro-optic modulator, which
  can coherently couple an optical cavity mode to a microwave
  resonator mode and enable novel quantum operations on the two modes,
  including laser cooling of the microwave mode, electro-optic
  entanglement, and backaction-evading optical measurement of a
  microwave quadrature. In this sequel, I focus on the quantum
  input-output relations between traveling optical and microwave
  fields coupled to a cavity electro-optic modulator. With
  red-sideband optical pumping, the relations are shown to resemble
  those of a beam splitter for the traveling fields, so that in the
  ideal case of zero parasitic loss and critical coupling, microwave
  photons can be coherently up-converted to ``flying'' optical photons
  with unit efficiency, and vice versa. With blue-sideband pumping,
  the modulator acts as a nondegenerate parametric amplifier, which
  can generate two-mode squeezing and hybrid entangled photon pairs at
  optical and microwave frequencies. These fundamental operations
  provide a potential bridge between circuit quantum electrodynamics
  and quantum optics.
\end{abstract}

\maketitle
\section{Introduction}

The rapid recent progress in circuit quantum electrodynamics (QED)
\cite{circuit_qed} has motivated the question of how superconducting
microwave circuits can be interfaced with quantum optics technology
for long-distance quantum information transfer.  This task requires
efficient and coherent frequency conversion between microwave and
optical photons. Existing proposals involve the use of mechanical
oscillators as mediators between electrical and optical systems
\cite{mechanics}, but a more straightforward way is to take advantage
of the well known Pockels electro-optic effect in a noncentrosymmetric
material, such as lithium niobate \cite{yariv}. The Pockels effect is
the change in the optical index of refraction of a material under an
applied voltage. A Pockels cell can be satisfactorily modeled as a
broadband second-order nonlinear optical medium and a capacitor on the
electrical side \cite{yariv}, so the effect is inherently coherent and
suitable for quantum optics experiments, much like the use of
second-order nonlinear crystals in optical parametric amplifiers and
oscillators.  In the classical regime, high-quality cavity
electro-optic modulators that can resonantly couple microwave and
optical fields have been extensively studied and experimentally
demonstrated \cite{cohen,ilchenko,matsko,savchenkov}, but a quantum
analysis of the photon frequency conversion problem is still lacking.

In the previous paper \cite{cqeo}, I have developed a quantum model of
cavity electro-optic modulators that can be used to address the
frequency conversion problem. While the previous paper focuses on the
analogy between electro-optics and optomechanics and the interactions
between resonator modes, the present paper studies the relations
between the traveling microwave and optical fields coupled to the
cavities and the conversion efficiencies in the presence of parasitic
losses. I consider two modes of operations: red-sideband optical
pumping and blue-sideband optical pumping. Red-sideband pumping in the
classical regime has been considered previously in
Refs.~\cite{savchenkov} with the assumption that the microwave field
is undepleted; here I shall do a quantum analysis assuming that the
optical pump is undepleted instead and allow the microwave fields and
the up-converted optical fields to exchange energy. This process is
shown to be a fundamentally noiseless operation resembling that of a
variable beam-splitter, so that in the ideal case of zero parasitic
loss and critical coupling, microwave photons can be coherently
converted to optical photons with unit efficiency, and vice
versa. With blue-sideband pumping, the electro-optic modulator acts as
a nondegenerate parametric amplifier, which can generate two-mode
squeezing and hybrid entangled photon pairs at optical and microwave
frequencies. Given the fundamental importance of beam-splitters and
parametric amplifiers in quantum optics \cite{mandel}, such operations
enabled by the cavity electro-optic modulator should be similarly
useful for future quantum optical interconnect technology, if the
technical challenges of implementing a quantum-efficient cavity
electro-optic modulator can be overcome.

\section{Model}

\begin{figure}[htbp]
\centerline{\includegraphics[width=0.48\textwidth]{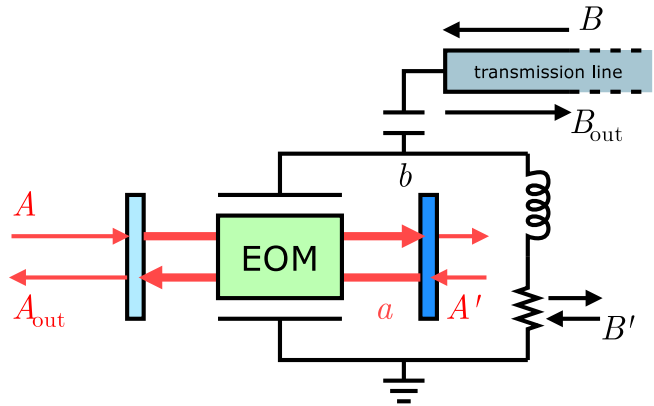}}
\caption{(Color online). Schematic of a cavity electro-optic modulator
  coupled to traveling fields. The physics remains essentially the
  same regardless of the actual types of the optical and microwave
  resonators.}
\label{oe_setup}
\end{figure}

As shown in Fig.~\ref{oe_setup}, the cavity electro-optic modulator
model considered here is a generalization of the one in
Ref.~\cite{cqeo} and also includes traveling optical and microwave
fields coupled to the optical and microwave resonators. $A$ and
$A_{\rm out}$ are the input and output optical field annihilation
operators, $B$ and $B_{\rm out}$ are the input and output microwave
field annihilation operators, $A'$ and $B'$ are the quantum Langevin
operators coupled through parasitic losses in the optical and
microwave resonators \cite{mandel}, and $a$ and $b$ are the optical
and microwave resonator-mode annihilation operators with resonance
frequencies $\omega_{a,b}$.  The relevant commutation relations are
\begin{align}
\Bk{A(t),A^\dagger(t')} &= \delta(t-t'),
\\
\Bk{B(t),B^\dagger(t')} &= \delta(t-t'),
\\
\Bk{A'(t),A'^\dagger(t')} &= \delta(t-t'),
\\
\Bk{B'(t),B'^\dagger(t')} &= \delta(t-t'),
\\
\Bk{a,a^\dagger} &= 1,
\\
\Bk{b,b^\dagger} &= 1.
\end{align}
In the following, I shall consider optical pumping at frequency
$\omega_a-\omega_b$ or $\omega_a+\omega_b$. Following the terminology
of optical parametric oscillators, I shall call the configuration
\emph{doubly resonant} (refering to the resonances at $\omega_a$ and
$\omega_b$) if the optical cavity is off-resonant at the pump
frequency and \emph{triply resonant} if the cavity is also resonant at
the pump frequency.

\section{Red-sideband optical pumping}

\subsection{Laplace analysis}

\begin{figure}[htbp]
\centerline{\includegraphics[width=0.8\textwidth]{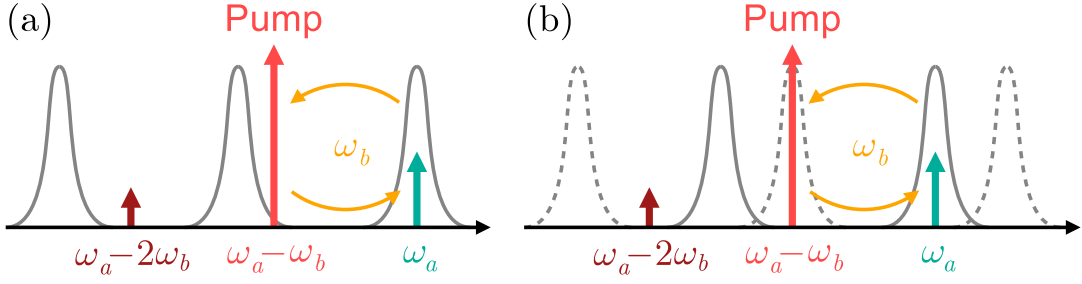}}
\caption{(Color online). Red-sideband optical pumping schemes. (a) a
  doubly-resonant configuration with an off-resonant pump. (b) a
  triply-resonant configuration with a resonant pump in a different
  polarization mode \cite{savchenkov}. Both schemes suppress
  interactions with the off-resonant field at $\omega_a-2\omega_b$.}
\label{red-sideband}
\end{figure}

Consider first red-sideband optical pumping at a frequency
$\omega_a-\omega_b$, as depicted in Fig.~\ref{red-sideband}. Assume
that the optical cavity is off-resonant at $\omega_a-2\omega_b$, so
that the interactions between the pump and the optical field at
$\omega_a-2\omega_b$ can be neglected.  This can be achieved for a
Fabry-P\'erot or whispering-gallery-mode cavity if $\omega_b$ does not
coincide with the free spectral range, so that the pump is
off-resonant in a doubly resonant configuration \cite{cqeo}, or if the
pump and the optical mode at $\omega_a$ are modes with different
polarizations in a triply resonant configuration
\cite{savchenkov}. The resulting equations of motion in an appropriate
rotating frame become
\begin{align}
\diff{a}{t} &= ig \alpha b -\frac{\Gamma_a}{2} a + 
\sqrt{\gamma_a}A + \sqrt{\gamma_a'}A',
\label{a}\\
\diff{b}{t} &= ig \alpha^* a -\frac{\Gamma_b}{2} b + 
\sqrt{\gamma_b}B + \sqrt{\gamma_b'}B',
\label{b}\\
A_{\rm out} &= \sqrt{\gamma_a} a - A,
\label{Aout}\\
B_{\rm out} &= \sqrt{\gamma_b} b - B.
\label{Bout}
\end{align}
where 
\begin{align}
g &\equiv \frac{\omega_a n^3 r l}{c\tau d}
\bk{\frac{\hbar\omega_b}{2C}}^{1/2}
\end{align}
is the electro-optic coupling coefficient in units of Hertz
\cite{cqeo}, $n$ is the optical index of refraction inside the
electro-optic medium, $r$ is the electro-optic coefficient in units of
m/V \cite{yariv}, $l$ is the length of the medium, $d$ is the
thickness, $\tau$ is the optical round-trip time, $C$ is the
capacitance of the microwave resonator, $\Gamma_{a,b}$ are the total
decay rates of the optical and microwave modes and are sums of the
traveling-field coupling rates $\gamma_{a,b}$ and parasitic decay
rates $\gamma_{a,b}'$, viz.,
\begin{align}
\Gamma_a &= \gamma_a+\gamma_a',
&
\Gamma_b &= \gamma_b + \gamma_b',
\end{align}
and $\alpha$ is the normalized pump field amplitude, such that
$|\alpha|^2$ is the number of pump photons inside the cavity.

\begin{figure}[htbp]
\centerline{\includegraphics[width=0.48\textwidth]{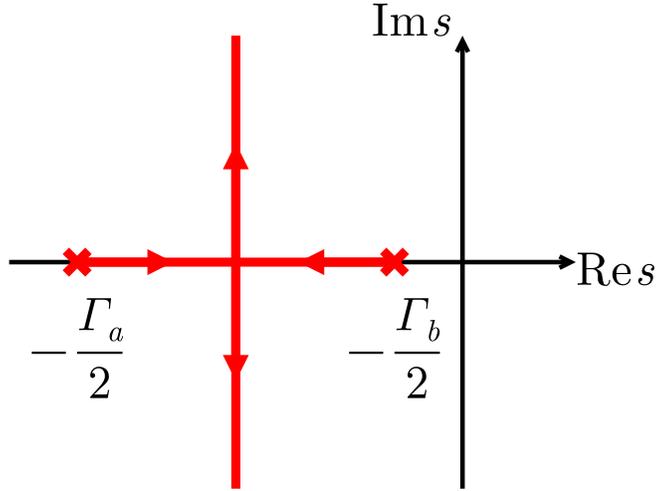}}
\caption{(Color online). Root-locus plot for increasing red-sideband
  pump strength $|g\alpha|$.}
\label{root_locus}
\end{figure}

Equations (\ref{a})-(\ref{Bout}) are most easily solved using the
Laplace transform, viz.,
\begin{align}
\tilde f(s) &\equiv \int_0^\infty dt f(t)\exp(-st),
\end{align}
so that, for example,
\begin{align}
\diff{a(t)}{t} &\to s\tilde a(s) - a(0).
\end{align}
The solutions for $A_{\rm out}$ and $B_{\rm out}$ in the Laplace
domain are given by
\begin{widetext}
\begin{align}
\bk{\begin{array}{c}\tilde A_{\rm out}(s)\\ \tilde B_{\rm out}(s)\end{array}}
&= \bk{\begin{array}{cc}F_{Aa}(s) & F_{Ab}(s) \\
F_{Ba}(s) & F_{Bb}(s)\end{array}}
\bk{\begin{array}{c}a(0)\\ b(0)\end{array}} 
+ \bk{\begin{array}{cccc}S_{AA}(s)&S_{AB}(s)&S_{AA'}(s) &S_{AB'}(s)\\
S_{BA}(s)&S_{BB}(s) &S_{BA'}(s)&S_{BB'}(s)\end{array}}
\bk{\begin{array}{c}\tilde A(s) \\\tilde B(s) \\ \tilde A'(s) \\\tilde B'(s)\end{array}}.
\label{matrix_io}
\end{align}
\end{widetext}
The first part of the solution that depends on an $F$ matrix, $a(0)$,
and $b(0)$ is the transient solution. Explicitly, the $F$ matrix is
given by
\begin{align}
F(s) &= \frac{1}{D(s)}\bk{\begin{array}{cc}\sqrt{\gamma_a}\bk{s+\frac{\Gamma_b}{2}}
& i\sqrt{\gamma_a}g\alpha\\
i\sqrt{\gamma_b}g\alpha^* & \sqrt{\gamma_b}\bk{s+\frac{\Gamma_a}{2}}
\end{array}},
\end{align}
where the denominator is
\begin{align}
D(s) &\equiv \bk{s+\frac{\Gamma_a}{2}}\bk{s+\frac{\Gamma_b}{2}}+|g\alpha|^2
\\
&= (s-p_+)(s-p_-).
\end{align}
The poles of the transfer functions $p_\pm$, given by
\begin{align}
p_\pm &\equiv -\frac{\Gamma_a+\Gamma_b}{4}\pm
\sqrt{\bk{\frac{\Gamma_a-\Gamma_b}{4}}^2-|g\alpha|^2},
\label{poles}
\end{align}
play a crucial role in the system dynamical response.
Figure~\ref{root_locus}, the so-called root-locus plot \cite{nise},
shows the loci of the poles on the complex plane as $|g\alpha|$ is
increased. This plot is typical of a damped harmonic oscillator. When
\begin{align}
|g\alpha| > \frac{|\Gamma_a-\Gamma_b|}{4},
\end{align}
the poles become complex, indicating a phenomenon analogous to Rabi
splitting \cite{mandel}. The coupled electro-optic response then
becomes oscillatory.

\subsection{Electro-optic beam-splitting}
While the transient solution can be relevant to the task of reading
out resonator modes, the $S$ matrix, which relates the traveling
fields, is of more interest to frequency conversion:
\begin{widetext}
\begin{align}
S(s) &= \frac{1}{D(s)}
\bk{\begin{array}{cccc}
\bk{-s+\frac{\gamma_a-\gamma_a'}{2}}
\bk{s+\frac{\Gamma_b}{2}}-|g\alpha|^2
&
ig\alpha\sqrt{\gamma_a\gamma_b}
&
\sqrt{\gamma_a\gamma_a'}\bk{s+\frac{\Gamma_b}{2}}
&
ig\alpha\sqrt{\gamma_a\gamma_b'}
\\
ig\alpha^*\sqrt{\gamma_a\gamma_b}
&
\bk{-s+\frac{\gamma_b-\gamma_b'}{2}}
\bk{s+\frac{\Gamma_a}{2}}-|g\alpha|^2
&
ig\alpha^*\sqrt{\gamma_a'\gamma_b}
&
\sqrt{\gamma_b\gamma_b'}\bk{s+\frac{\Gamma_a}{2}}
\end{array}}.
\label{Smatrix}
\end{align}
The spectral behavior of the system is obtained by neglecting the
transient solution and substituting $s = -i\omega$ in the $S$ matrix,
where $\omega$ is the detuning with respect to the carrier frequencies
$\omega_{a,b}$. The Fourier transforms of the input and output fields
are related by the $S(-i\omega)$ matrix:
\begin{align}
\hat f(\omega) &\equiv \intall dt f(t)\exp(i\omega t),
\\
\bk{\begin{array}{c}\hat A_{\rm out}(\omega)\\ \hat B_{\rm out}(\omega)\end{array}}
&=  \bk{\begin{array}{cccc}S_{AA}(-i\omega)&S_{AB}(-i\omega)&S_{AA'}(-i\omega) &S_{AB'}(-i\omega)\\
S_{BA}(-i\omega)&S_{BB}(-i\omega) &S_{BA'}(-i\omega)&S_{BB'}(-i\omega)\end{array}}
\bk{\begin{array}{c}\hat A(\omega) \\\hat B(\omega) \\ \hat A'(\omega) \\\hat B'(\omega)\end{array}}.
\label{matrix_io_freq}
\end{align}
\end{widetext}
Equation~(\ref{matrix_io_freq}) then resembles the spectral-domain
input-output relations for a lossy beam splitter with quantum Langevin
noise fields $\hat A'$ and $\hat B'$ \cite{barnett}.

For frequency conversion, the most important quantity is the
electro-optic conversion efficiency, defined by
\begin{align}
R(\omega) &\equiv |S_{AB}(-i\omega)|^2 = |S_{BA}(-i\omega)|^2 
\\
&= \frac{|g\alpha|^2\gamma_a\gamma_b}
{|(-i\omega-p_+)(-i\omega-p_-)|^2}.
\label{efficiency}
\end{align}
At zero detuning ($\omega=0$),
\begin{align}
R(0) &= \frac{4\eta G_0}{(1+G_0)^2},
\end{align}
where
\begin{align}
G_0 &\equiv \frac{4|g\alpha|^2}{\Gamma_a\Gamma_b}
\end{align}
is analogous to the \emph{cooperativity parameter} in cavity QED
\cite{kimble} and
\begin{align}
\eta &\equiv \frac{\gamma_a\gamma_b}{\Gamma_a\Gamma_b}
\end{align}
is the intrinsic efficiency of the system. 

\begin{figure}[htbp]
\centerline{\includegraphics[width=0.48\textwidth]{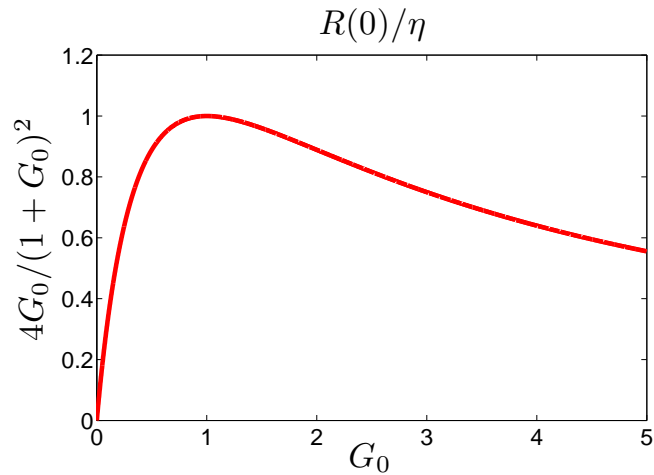}}
\caption{(Color online). Conversion efficiency $R(0)/\eta$ at zero
  detuning versus the cooperativity parameter $G_0$.}
\label{conversion}
\end{figure}

\begin{figure}[htbp]
\centerline{\includegraphics[width=0.48\textwidth]{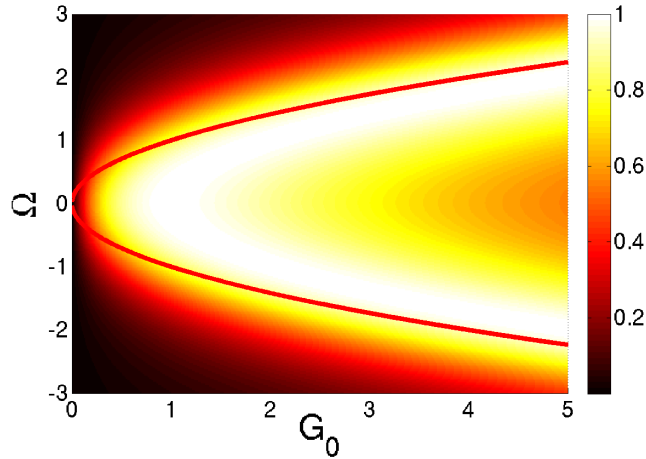}}
\caption{(Color online). The color plot shows the conversion
  efficiency $R(\omega)/\eta$ at $\Gamma_a=\Gamma_b$ and a fixed
  $\eta$ with respect to $G_0$ on the horizontal axis and
  $\Omega\equiv 2\omega/\sqrt{\Gamma_a\Gamma_b}$ on the vertical
  axis. The solid lines are the imaginary parts of the poles.}
\label{bandwidth_G}
\end{figure}

\begin{figure}[htbp]
\centerline{\includegraphics[width=0.48\textwidth]{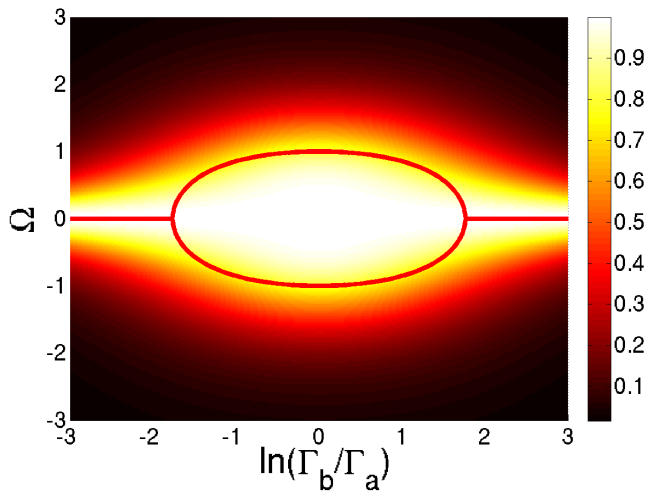}}
\caption{(Color online). The color plot shows the conversion
  efficiency $R(\omega)/\eta$ at critical coupling ($G_0=1$) and a fixed
  $\eta$ with respect to $\ln(\Gamma_b/\Gamma_a)$ on the
  horizontal axis and $\Omega\equiv 2\omega/\sqrt{\Gamma_a\Gamma_b}$
  on the vertical axis.  The solid lines are the imaginary parts of
  the poles. The bandwidth is maximum when $\Gamma_a = \Gamma_b$ and
  the imaginary parts of the poles are the farthest apart.}
\label{bandwidth}
\end{figure}

Figure~\ref{conversion} plots the conversion efficiency $R(0)/\eta$ at
zero detuning.  The highest efficiency at zero detuning is achieved
when
\begin{align}
G_0 &= 1, & R(0) &= \eta.
\end{align}
Since the zero-detuning efficiency drops when $G_0 > 1$, $G_0 = 1$ can be
regarded as a \emph{critical coupling} condition. For other
frequencies, the efficiency given by Eq.~(\ref{efficiency}) depends on
the product of the distances between $-i\omega$ and the poles $p_\pm$
on the complex plane. Figure~\ref{bandwidth_G} plots the conversion
efficiency with respect to the normalized detuning frequency
$\Omega\equiv 2\omega/\sqrt{\Gamma_a\Gamma_b}$ and increasing $G_0$,
showing that the highest efficiencies indeed occur at frequencies near
the poles. The conversion bandwidth is thus maximum when the imaginary
parts of the poles are the farthest apart.  For a fixed $|g\alpha|^2$,
this means that
\begin{align}
\Gamma_a &= \Gamma_b,
\end{align}
and the resonators should ideally have the same decay rates.
Figure~\ref{bandwidth}, which plots the efficiency at critical
coupling against $\ln (\Gamma_b/\Gamma_a)$, confirms this behavior.

In the case of $\gamma_{a,b}' = 0$, $S_{AA'}$, $S_{AB'}$, $S_{BA'}$,
and $S_{BB'}$ are all zero, and the ideal lossless beam-splitting
relations are recovered:
\begin{align}
\bk{\begin{array}{c}\hat A_{\rm out}(\omega)\\ \hat B_{\rm out}(\omega)\end{array}}
&= \bk{\begin{array}{cc}S_{AA}(-i\omega)&S_{AB}(-i\omega) \\
S_{BA}(-i\omega)&S_{BB}(-i\omega) \end{array}}
\bk{\begin{array}{c}\hat A(\omega)\\ \hat B(\omega) \end{array}},
\end{align}
in which case the conversion efficiency at zero detuning can be
perfect at critical coupling:
\begin{align}
G_0 &= 1, & R(0) &= 1,
&
T(0) &\equiv |S_{AA}(0)|^2 = |S_{BB}(0)|^2 = 0.
\end{align}
Faithful frequency conversion thus requires relatively low parasitic
losses ($\gamma_{a}' \ll \gamma_{a}, \gamma_{b}' \ll \gamma_b$) and
the critical coupling condition ($G_0 = 1$).

\section{Blue-sideband optical pumping}

\subsection{Laplace analysis}

\begin{figure}[htbp]
\centerline{\includegraphics[width=0.8\textwidth]{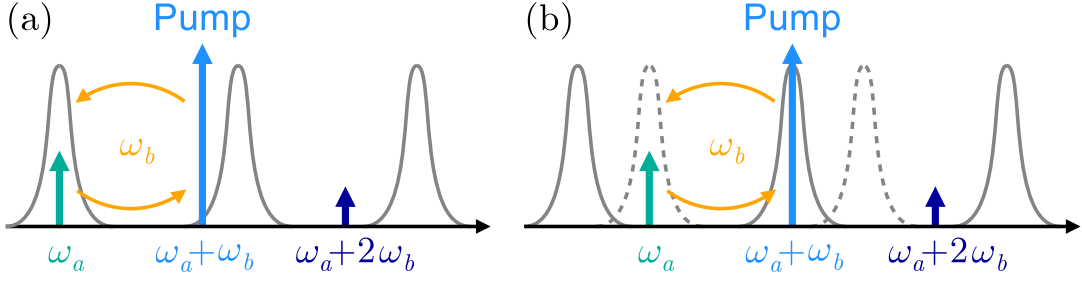}}
\caption{(Color online). Blue-sideband optical pumping scheme in (a) a
  doubly-resonant configuration with an off-resonant pump and (b) a
  triply-resonant configuration with a resonant pump.}
\label{blue-sideband}
\end{figure}

\begin{figure}[htbp]
\centerline{\includegraphics[width=0.48\textwidth]{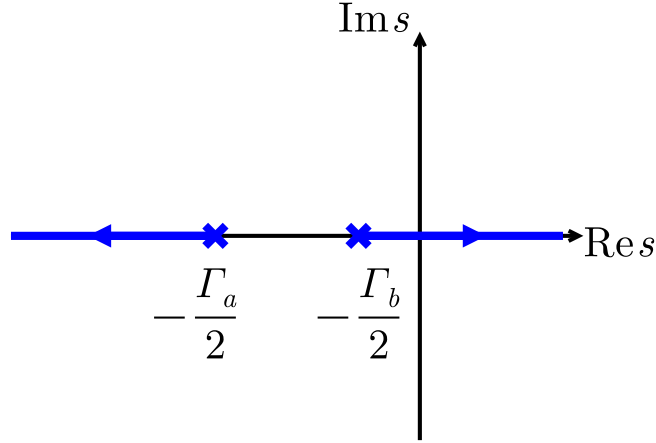}}
\caption{(Color online). Root-locus plot for increasing blue-sideband
  pump strength $|g\alpha|$.}
\label{root_locus_blue}
\end{figure}

The analysis of a blue-sideband optical pumping scheme
(Fig.~\ref{blue-sideband}) is similar; the equations of motion
are now given by
\begin{align}
\diff{a}{t} &= ig \alpha b^\dagger -\frac{\Gamma_a}{2} a + \sqrt{\gamma_a}A + \sqrt{\gamma_a'}A',
\\
\diff{b}{t} &= ig \alpha a^\dagger -\frac{\Gamma_b}{2} b + \sqrt{\gamma_b}B + \sqrt{\gamma_b'}B',
\\
A_{\rm out} &= \sqrt{\gamma_a} a - A,
\\
B_{\rm out} &= \sqrt{\gamma_b} b - B.
\end{align}
The solutions for $A_{\rm out}$ and $B_{\rm out}^\dagger$ in the Laplace domain
can be written as
\begin{widetext}
\begin{align}
\bk{\begin{array}{c}\tilde A_{\rm out}(s)\\ \tilde B_{\rm out}^\dagger(s^*)\end{array}}
&= \bk{\begin{array}{cc}\mathcal F_{Aa}(s) & \mathcal F_{Ab}(s) \\
\mathcal F_{Ba}(s) & \mathcal F_{Bb}(s)\end{array}}
\bk{\begin{array}{c}a(0)\\ b^\dagger(0)\end{array}} 
+ \bk{\begin{array}{cccc}\mathcal S_{AA}(s)&
\mathcal S_{AB}(s) &\mathcal S_{AA'}(s)&\mathcal S_{AB'}(s)\\
\mathcal S_{BA}(s)&\mathcal S_{BB}(s) &\mathcal S_{BA'}(s)&\mathcal S_{BB'}(s)\end{array}}
\bk{\begin{array}{c}\tilde A(s)  \\\tilde B^\dagger(s^*)\\\tilde A'(s)
\\ \tilde B'^\dagger(s^*)\end{array}}.
\label{matrix_io_blue}
\end{align}
\end{widetext}
These relations suggest that the electro-optic modulator now acts as a
nondegenerate parametric amplifier. The $\mathcal F$ matrix is
\begin{align}
\mathcal F(s) &= \frac{1}{\mathcal D(s)}
\bk{\begin{array}{cc}\sqrt{\gamma_a}(s+\Gamma_b/2)& i\sqrt{\gamma_a}g\alpha\\
-i\sqrt{\gamma_a}g\alpha^* &\sqrt{\gamma_b}(s+\Gamma_a/2)
\end{array}},
\\
\mathcal D(s) &\equiv \bk{s+\frac{\Gamma_a}{2}}\bk{s+\frac{\Gamma_b}{2}}-|g\alpha|^2
\\
&=(s-\pi_+)(s-\pi_-).
\end{align}
The poles are
\begin{align}
\pi_\pm &= -\frac{\Gamma_a+\Gamma_b}{4}\pm
\sqrt{\bk{\frac{\Gamma_a-\Gamma_b}{4}}^2+|g\alpha|^2},
\end{align}
which, as shown in Fig.~\ref{root_locus_blue}, follow very different
loci than the ones for red-sideband pumping in Fig.~\ref{root_locus}
and remain real. When
\begin{align}
G_0 &\equiv \frac{4|g\alpha|^2}{\Gamma_a\Gamma_b} \ge 1,
&
\pi_+ &\ge 0,
\end{align}
the $\pi_+$ pole moves to the right-half plane, and the system becomes
unstable. In other words, $G_0 \ge 1$ is the threshold condition for
electro-optic parametric \emph{oscillation}.

\subsection{Electro-optic parametric amplification}
\begin{figure}[htbp]
\centerline{\includegraphics[width=0.48\textwidth]{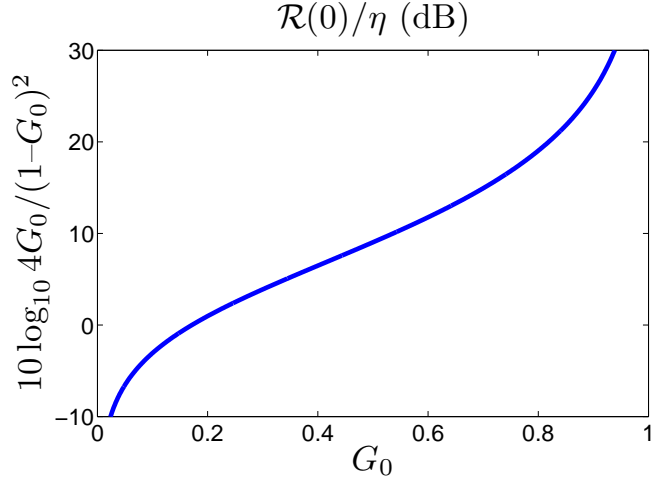}}
\caption{(Color online). Idler gain $\mathcal R(0)/\eta$ in dB at zero
  detuning versus $G_0$.}
\label{idler_gain}
\end{figure}

Below threshold ($G_0 < 1$), the input-output relations for the
nondegenerate parametric amplifier are
\begin{widetext}
\begin{align}
\mathcal S(s) &= \frac{1}{\mathcal D(s)}
\bk{\begin{array}{cccc}
\bk{-s+\frac{\gamma_a-\gamma_a'}{2}}
\bk{s+\frac{\Gamma_b}{2}}+|g\alpha|^2
&
ig\alpha\sqrt{\gamma_a\gamma_b}
&
\sqrt{\gamma_a\gamma_a'}\bk{s+\frac{\Gamma_b}{2}}
&
ig\alpha\sqrt{\gamma_a\gamma_b'}
\\
-ig\alpha^*\sqrt{\gamma_a\gamma_b}
&
\bk{-s+\frac{\gamma_b-\gamma_b'}{2}}
\bk{s+\frac{\Gamma_a}{2}}+|g\alpha|^2
&
-ig\alpha^*\sqrt{\gamma_a'\gamma_b}
&
\sqrt{\gamma_b\gamma_b'}\bk{s+\frac{\Gamma_a}{2}}
\end{array}}.
\end{align}
The parametric gains in the spectral domain are given by
\begin{align}
\bk{\begin{array}{c}\hat A_{\rm out}(\omega)\\ \hat B_{\rm out}^\dagger(-\omega)\end{array}}
&=  \bk{\begin{array}{cccc}\mathcal S_{AA}(-i\omega)&\mathcal S_{AB}(-i\omega)
&\mathcal S_{AA'}(-i\omega) &\mathcal S_{AB'}(-i\omega)\\
\mathcal S_{BA}(-i\omega)&\mathcal S_{BB}(-i\omega) &\mathcal S_{BA'}(-i\omega)
&\mathcal S_{BB'}(-i\omega)\end{array}}
\bk{\begin{array}{c}\hat A(\omega) \\\hat B^\dagger(-\omega) \\ 
\hat A'(\omega) \\\hat B'^\dagger(-\omega)\end{array}}.
\end{align}
\end{widetext}
In particular, the amplified electro-optic conversion efficiency, or
the idler gain, is
\begin{align}
\mathcal R(\omega) &\equiv |\mathcal S_{BA}(-i\omega)|^2 = |\mathcal S_{AB}(-i\omega)|^2
\\
&= \frac{|g\alpha|^2\gamma_a\gamma_b}{(\omega^2+\pi_+^2)(\omega^2+\pi_-^2)}.
\end{align}
With the real poles, the spectral behavior of the amplifier in general
resembles a bandpass filter around zero detuning, at which the gain is
\begin{align}
\mathcal R(0) &=  \frac{4\eta G_0}{(1-G_0)^2}.
\end{align}
Figure~\ref{idler_gain} plots this function in dB against the
cooperativity parameter $G_0$. Unlike the conversion efficiency for
red-sideband pumping in Fig.~\ref{conversion}, the gain increases
indefinitely for increasing $G_0$ until the threshold condition.

\begin{figure}[htbp]
\centerline{\includegraphics[width=0.48\textwidth]{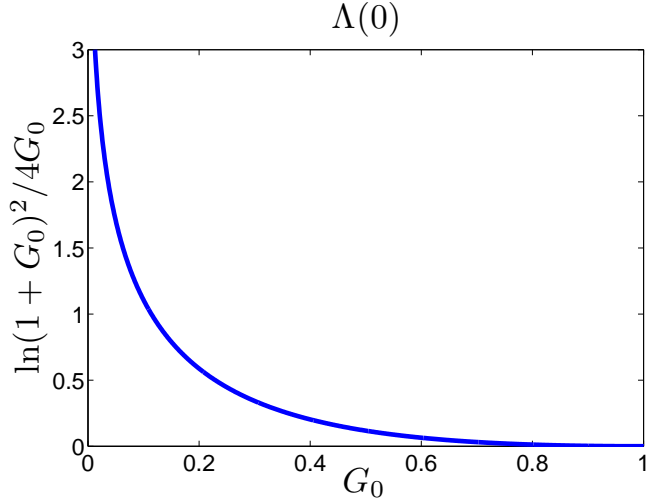}}
\caption{(Color online). Nonclassicality parameter $\Lambda(0)$ at
  zero detuning versus $G_0$.}
\label{nonclassical}
\end{figure}

Parametric amplification may be useful for electro-optic conversion in
the classical regime, but amplification in the quantum regime
necessarily comes with noise. For coherent-state inputs, the noise
statistics are completely determined by
\begin{align}
\Avg{\hat A_{\rm out}^\dagger(\omega)\hat A_{\rm out}(\omega')}
&= 
\Avg{\hat A_{\rm out}^\dagger(\omega)}\Avg{\hat A_{\rm out}(\omega')}
+2\pi\delta(\omega-\omega')
\Bk{\mathcal R(\omega) + \mathcal R_A'(\omega)},
\label{covariance_A}
\\
\Avg{\hat B_{\rm out}^\dagger(\omega)\hat B_{\rm out}(\omega')}
&= 
\Avg{\hat B_{\rm out}^\dagger(\omega)}\Avg{\hat B_{\rm out}(\omega')}
+2\pi\delta(\omega-\omega')
\Bk{\mathcal R(\omega) + \mathcal R_B'(\omega)},
\label{covariance_B}
\\
\Avg{\hat A_{\rm out}(\omega)\hat B_{\rm out}(\omega')}
&= \Avg{\hat A_{\rm out}(\omega)}\Avg{\hat B_{\rm out}(\omega')}
+2\pi\delta(\omega+\omega')\mathcal K(\omega),
\end{align}
where
\begin{align}
\mathcal R_A'(\omega) &\equiv |\mathcal S_{AB'}(-i\omega)|^2
= \frac{|g\alpha|^2\gamma_a\gamma_b'}{(\omega^2+\pi_+^2)(\omega^2+\pi_-^2)},
\\
\mathcal R_B'(\omega) &\equiv |\mathcal S_{BA'}(-i\omega)|^2
= \frac{|g\alpha|^2\gamma_a'\gamma_b}{(\omega^2+\pi_+^2)(\omega^2+\pi_-^2)},
\\
\mathcal K(\omega)
&\equiv
\mathcal S_{AA}(-i\omega)\mathcal S_{BA}^*(-i\omega)
+\mathcal S_{AA'}(-i\omega)\mathcal S_{BA'}^*(-i\omega)
\\
&=
\frac{ig\alpha\sqrt{\gamma_a\gamma_b}}{(\omega^2+\pi_+^2)(\omega^2+\pi_-^2)}
\Bk{\bk{i\omega+\frac{\Gamma_a}{2}}\bk{-i\omega+\frac{\Gamma_b}{2}}
+|g\alpha|^2}.
\end{align}
To investigate the nonclassicality of the hybrid squeezed state when
the inputs are vacuum, one can use the optical equivalence theorem
\cite{mandel} to write the phase-sensitive covariance as
\begin{align}
\Avg{\hat A_{\rm out}(\omega)\hat B_{\rm out}(\omega')} &=
 \int D\mathcal A D\mathcal B
P_{\rm out}[\mathcal A,\mathcal B]
\mathcal A(\omega) \mathcal B(\omega'),
\end{align}
where $\mathcal A$ and $\mathcal B$ are classical fields and $P_{\rm
  out}[\mathcal A,\mathcal B]$ is the $P$ functional for the output
fields. If the $P$ representation is nonnegative, the Cauchy-Schwarz
inequality gives
\begin{align}
\abs{\Avg{\hat A_{\rm out}(\omega)\hat B_{\rm out}(\omega')}}^2&\le 
\int D\mathcal A D\mathcal B P_{\rm out}[\mathcal A,\mathcal B] |\mathcal A(\omega)|^2
\int D\mathcal A D\mathcal B P_{\rm out}[\mathcal A,\mathcal B] |\mathcal B(\omega')|^2
\\
&= \Avg{\hat A^\dagger_{\rm out}(\omega)\hat A_{\rm out}(\omega)}
\Avg{\hat B^\dagger_{\rm out}(\omega')\hat B_{\rm out}(\omega')}.
\end{align}
This implies that, for a classical state,
\begin{align}
|\mathcal K(\omega)|^2 &= \frac{1}{\eta}\mathcal R^2(\omega)+\mathcal R(\omega)
\\
&\le \Bk{\mathcal R(\omega)+\mathcal R_A'(\omega)}
\Bk{\mathcal R(\omega)+\mathcal R_B'(\omega)}
\\
&= \frac{1}{\eta}\mathcal R^2(\omega)
\equiv |\mathcal K_c(\omega)|^2.
\end{align}
One can then define a nonclassicality parameter as
\begin{align}
\Lambda(\omega) &\equiv
\ln \frac{|\mathcal K(\omega)|^2}{|\mathcal K_c(\omega)|^2}
=
\ln \Bk{1+\frac{\eta}{\mathcal R(\omega)}}.
\end{align}
At zero detuning,
\begin{align}
\Lambda(0) &= \ln \frac{(1+G_0)^2}{4G_0},
\end{align}
which depends on $G_0$ but not $\eta$. The phase-sensitive correlation
is strongly nonclassical ($\Lambda \gg 1$) when $G_0 \ll 1$ but
vanishes at threshold, as shown in Fig.~\ref{nonclassical}.  The
nonclassical correlation may be useful for quantum illumination
\cite{lloyd,tan} in the microwave regime with a retained optical
idler. Another electro-optic parametric amplifier may be used as a
receiver that combines the return microwave signal and the optical
idler and counts the optical photon number \cite{guha} to achieve
quantum-enhanced target detection \cite{lloyd,tan} and secure
communication \cite{shapiro_qi}.

In the case of $\gamma_{a,b}' = 0$, the ideal parametric-amplification
relations are recovered:
\begin{align}
\bk{\begin{array}{c}\hat A_{\rm out}(\omega)\\ 
\hat B_{\rm out}^\dagger(-\omega)\end{array}}
&= \bk{\begin{array}{cc}\mathcal S_{AA}(-i\omega)&\mathcal S_{AB}(-i\omega) \\
\mathcal S_{BA}(-i\omega)&\mathcal S_{BB}(-i\omega) \end{array}}
\bk{\begin{array}{c}\hat A(\omega)\\ \hat B^\dagger(-\omega) \end{array}}.
\end{align}
The standard analysis of two-mode parametric amplification and
squeezing \cite{shapiro,caves} then applies.

\subsection{Hybrid entangled photons}
An alternative way of studying the entanglement between the two fields
is to consider the Schr\"odinger picture. Assume that the idler gain
is small enough such that one can write
\begin{align}
\hat A_{\rm out}(\omega) &\approx \hat A(\omega)-i\Bk{\hat A(\omega),\epsilon},
\end{align}
$\epsilon$ being an Hermitian operator, and likewise for the other
output fields.  It is not difficult to show that this can be satisfied
if $G_0 \ll 1$ and
\begin{widetext}
\begin{align}
\epsilon &= \intall \frac{d\omega}{2\pi} 
\Bk{i\mathcal S_{AB}(-i\omega)\hat A^\dagger(\omega)
\hat B^\dagger(-\omega)+i\mathcal S_{AB'}(-i\omega)\hat A^\dagger(\omega)\hat B'^\dagger(-\omega)
+i\mathcal S_{BA'}^*(-i\omega)\hat A'^\dagger(\omega)\hat B^\dagger(-\omega)
+\textrm{H.c.}},
\end{align}
H.c.\ denoting the Hermitian conjugate.  One can then
write the unitary evolution operator as
\begin{align}
U \approx 1-i\epsilon,
\end{align}
and the Schr\"odinger-picture output state for a vacuum input
state $\ket{\textrm{vac}}$ as
\begin{align}
\ket{\Psi} &=U\ket{\textrm{vac}} \approx (1-i\epsilon)\ket{\textrm{vac}}
\\
&= \ket{\textrm{vac}} +\intall\frac{d\omega}{2\pi}
\Bk{\mathcal S_{AB}(-i\omega)\hat A^\dagger(\omega)
\hat B^\dagger(-\omega)+\mathcal S_{AB'}(-i\omega)\hat A^\dagger(\omega)\hat B'^\dagger(-\omega)
+\mathcal S_{BA'}^*(-i\omega)\hat A'^\dagger(\omega)\hat B^\dagger(-\omega)}\ket{\textrm{vac}}.
\end{align}
\end{widetext}
Tracing out the inaccessible $A'$ and $B'$ modes and denoting the
vacuum state in the subspace of $A$ and $B$ modes as $\ket{0,0}$, one
obtains
\begin{align}
\rho_{AB} &=\trace_{A'B'} \ket{\Psi}\bra{\Psi}
\\
&\approx \ket{\psi}\bra{\psi}+\intall\frac{d\omega}{2\pi}
\mathcal R_A'(\omega)
\ket{1_\omega,0}\bra{1_\omega,0}
+\intall \frac{d\omega}{2\pi}
\mathcal R_B'(\omega)\ket{0,1_\omega}\bra{0,1_\omega},
\\
\ket{\psi} &\equiv \ket{0,0}+\intall \frac{d\omega}{2\pi} \mathcal S_{AB}(-i\omega)
\ket{1_\omega,1_{-\omega}},
\end{align}
where the unnormalized Fock states are defined by
\begin{align}
\ket{1_\omega,0} &\equiv \hat A^\dagger(\omega)\ket{0,0},
\\
\ket{0,1_\omega} &\equiv \hat B^\dagger(\omega)\ket{0,0},
\\
\ket{1_\omega,1_{-\omega}} &\equiv \hat A^\dagger(\omega)
\hat B^\dagger(-\omega)\ket{0,0}.
\end{align}
Thus, $\mathcal R(\omega)$ is the entangled photon-pair generation
rate per Hertz and $\mathcal R_{A,B}'(\omega)$ are the accidental
photon generation rates per Hertz. If, for instance, an optical photon
is used to herald a microwave photon, the heralding efficiency is
\begin{align}
\frac{\mathcal R(\omega)}{\mathcal R_{A}'(\omega)+\mathcal R(\omega)}
&= \frac{\gamma_{b}}{\Gamma_{b}},
\end{align}
which suggests that $\gamma_a' \ll \gamma_a$ and $\gamma_b' \ll
\gamma_b$ are desirable for generating pure entangled photons. The
entangled photons are frequency-anticorrelated, as one would expect
from energy conservation.

\section{Conclusion}
The most important result of this paper is that efficient
electro-optic frequency conversion requires the cooperativity paramter
$G_0$ and the intrinsic efficiency $\eta$ to be close to $1$. While it
should be possible to make $\eta$ close to $1$ using current microwave
and optical resonator technology, the $G_0$ of existing devices
\cite{ilchenko} is unfortunately on the order of $10^{-5}$ only
\cite{cqeo}. This should be enough for demonstrating hybrid entangled
photons, if the electro-optic modulator is kept at a cryogenic
temperature such that thermal microwave noise can be neglected, but
the small $G_0$ is hardly useful for coherent frequency
conversion. That said, given the potential room for improvement
\cite{cqeo}, which can make $g \sim 2\pi\times 5$~kHz and $G_0\sim 5$
for achievable parameters, and the head start enjoyed by
electro-optics technology \cite{cohen,ilchenko,savchenkov} in
experimental progress compared to competing electro-optomechanics
proposals \cite{mechanics}, which at this stage remain purely
theoretical, one should remain cautiously optimistic about the future
of quantum electro-optics.

\section*{Acknowledgments}
Discussions with Jeffrey Shapiro, Carlton Caves, Aaron Danner, Olivier
Pfister, Keith Schwab, Jun Ye, and James Thompson are gratefully
acknowledged. This material is based on work supported in
part by the Singapore National Research Foundation under NRF Award
No.~NRF-NRFF2011-07, NSF Grants No.~PHY-0903953, and No.~PHY-1005540.


\begin{thebibliography}{}
\bibitem{circuit_qed}
A.~Wallraff \textit{et al.},
Nature (London) \textbf{431}, 162 (2004);
A.~Blais \textit{et al.},
\pra \textbf{69}, 062320 (2004);
M.~H.~Devoret and J.~M.~Martinis,
Quant.\ Inform.\ Process.\ \textbf{3}, 163 (2004), and references therein;
J.~Clarke and F.~K.~Wilhelm,
Nature (London) \textbf{453}, 1031 (2008), and references therein.

\bibitem{mechanics}
L.~Tian and H.~Wang,
\pra \textbf{82}, 053806 (2010);
A.~H.~Safavi-Naeini and O.~J.~Painter,
New J.\ Phys.\ \textbf{13}, 013017 (2011);
C.~A.~Regal and K.~W.~Lehnert,
J.\ Phys.: Conf.\ Series \textbf{264}, 012025 (2011).

\bibitem{yariv}A.\ Yariv,
\textit{Quantum Electronics}
(Wiley, New York, 1989).

\bibitem{cohen}D.~A.~Cohen, M.~Hossein-Zadeh, and A.~F.~J.~Levi,
Electron.\ Lett.\ \textbf{37}, 300 (2001);
Solid-State Electron.\ \textbf{45}, 1577 (2001);
D.~A.~Cohen and A.~F.~J.~Levi, 
Electron. Lett. \textbf{37}, 37 (2001);
Solid State Electron.\ \textbf{45}, 495 (2001).

\bibitem{ilchenko}V.\ S.\ Ilchenko \textit{et al.},
\josab \textbf{20}, 333 (2003).

\bibitem{matsko}A.\ B.\ Matsko \textit{et al.},
Opt.\ Express \textbf{15}, 17401 (2007).

\bibitem{savchenkov}A.~A.~Savchenkov \textit{et al.},
\ol \textbf{34}, 1300 (2009);
IEEE Trans.\ Microw.\ Theor.\ Tech.\ \textbf{58}, 3167 (2010).

\bibitem{cqeo}M.~Tsang,
\pra \textbf{81}, 063837 (2010).

\bibitem{mandel}L.\ Mandel and E.\ Wolf,
\textit{Optical Coherence and Quantum Optics}
(Cambridge University Press, Cambridge, 1995).


\bibitem{nise}N.~S.~Nise,
\textit{Control Systems Engineering}
(Wiley, New York, 2011).

\bibitem{barnett}S.~M.~Barnett \textit{et al.},
\pra \textbf{57}, 2134 (1998).

\bibitem{kimble}H.~J.~Kimble,
Physica Scripta \textbf{T76}, 127 (1998).

\bibitem{lloyd}S.~Lloyd,
Science \textbf{321}, 1463 (2008).

\bibitem{tan}S.-H.~Tan \textit{et al.},
\prl \textbf{101}, 253601 (2008).

\bibitem{guha}S.~Guha and B.~I.~Erkmen,
\pra \textbf{80}, 052310 (2009).

\bibitem{shapiro_qi}J.~H.~Shapiro,
\pra \textbf{80}, 022320 (2009).

\bibitem{shapiro}
J.~H.~Shapiro and K.~X.~Sun,
\josab \textbf{11}, 1130 (1994);
J.~H.~Shapiro, 
Proc.\ SPIE \textbf{5111}, 382 (2003).

\bibitem{caves}C.~M.~Caves and B.~L.~Schumaker,
\pra \textbf{31}, 3068 (1985);
B.~L.~Schumaker and C.~M.~Caves,
\textit{ibid.}\ \textbf{31}, 3093 (1985).

\end{thebibliography}
\end{document}